\documentclass[12pt,letterpaper]{article}
\usepackage{jheppub}

\usepackage{graphicx}
\usepackage{bbm}
\usepackage{amsmath}
\usepackage{amssymb}
\usepackage{mathtools}
\usepackage{amsthm}

\usepackage[labelformat=simple]{subcaption}

\def\be{\begin{equation}}
\def\ee{\end{equation}}

\newtheoremstyle{named}{0.75\baselineskip}{0.75\baselineskip}{\itshape}{}{\bfseries}{.}{.5em}{#3}
\theoremstyle{named}
\newtheorem*{namedconjecture}{Conjecture}

\begin{document}

\begin{titlepage}


\setcounter{page}{1} \baselineskip=15.5pt \thispagestyle{empty}

\bigskip\

\vspace{2cm}
\begin{center}
{\LARGE \bfseries An Introduction to the \\ \vspace{.3cm} Weak Gravity Conjecture}

 \end{center}
\vspace{0.5cm}

\begin{center}
{\fontsize{14}{30}\selectfont Tom Rudelius}
\end{center}

\begin{center}
\vspace{0.25 cm}
\textsl{Department of Mathematical Sciences, Durham University, Durham, DH1 3LE, UK}

\vspace{0.25cm}

\end{center}
\vspace{1cm}
\noindent

The Weak Gravity Conjecture holds that gravity must be the weakest force. This is true of the familiar forces in our own universe---electromagnetism, for instance, is many orders of magnitude stronger than gravity. But the bold claim of the Weak Gravity Conjecture is that this statement is true, not only for electromagnetism in our universe, but for \emph{any} similar force in \emph{any} consistent universe governed by quantum mechanics. In this brief introduction, aimed at advanced undergraduates or beginning graduate students, we elaborate on the precise definition of the Weak Gravity Conjecture, the evidence for it, and some of the remarkable implications of it. We explain how the Weak Gravity Conjecture may play a role in bridging the gulf between the formal mathematics of string theory and the real-world observations of particle physics and cosmology.

\vspace{.9cm}

\bigskip
\noindent\today

\vspace{2.5cm}
\bigskip
\noindent Published in \emph{Contemporary Physics}.

\end{titlepage}

\setcounter{tocdepth}{2}

\hrule
\tableofcontents

\bigskip\medskip
\hrule
\bigskip\bigskip

\section{Invitation: The Weak Gravity Conjecture and the Electron \label{s.invitation}}

In a first course on Newtonian mechanics, we learn that the gravitational force between a pair of objects of mass $m_1$ and $m_2$ is given by
\begin{align}
    \vec F_{\rm grav} = - \frac{G m_1 m_2}{r^2} \hat r\,,
    \label{graveq}
\end{align}
where $G = 6.67 \times 10^{-11} \frac{\text{N} \cdot \text{m}}{\text{kg}^2}$ is Newton's gravitational constant.

Not long afterward, in a first course on electromagnetism, we learn that the electromagnetic force between a pair of objects of charge $q_1$ and $q_2$ is given by
\begin{align}
    \vec F_{\rm EM} =  \frac{k q_1 q_2}{r^2} \hat r\,,
\end{align}
where $k = 8.99 \times 10^9 \frac{\text{N} \cdot \text{m}}{\text{C}^2}$ is Coulomb's constant.

Let us consider now a situation in which a pair of electrons are separated by a large distance $r$.
Both electrons have a mass of $m_e = 9.11 \times 10^{-31}$ kg and a charge of $q_e = -1.60 \times 10^{-19}$ C, which means that the gravitational and electromagnetic forces felt by one electron due to its interaction with the other electron are given respectively by
\begin{align}
    \vec F_{\rm grav} = - \frac{G m_e^2}{r^2} \hat r = - \frac{5.54 \times 10^{-71} N \cdot m^2}{r^2} \hat r \\
    \vec F_{\rm EM} =  \frac{k q_e^2}{r^2} \hat r =  \frac{2.30 \times 10^{-28} N \cdot m^2}{r^2} \hat r \,.
\end{align}
In particular, the electromagnetic repulsion between the electrons is much stronger than their gravitational attraction, by around 43 orders of magnitude: in the absence of other forces, a pair of distantly separated electrons will accelerate in opposite directions. Thus, as far as the electron is concerned, gravity is the weakest force.

The idea behind the Weak Gravity Conjecture (WGC) \cite{Arkanihamed:2006dz} is to promote this observed fact of nature to a universal principle: any repulsive force, in any mathematically consistent universe,\footnote{The precise meaning of ``mathematically consistent'' will explained in further detail below.} must be stronger than gravity, in the sense that the charge of a particle is larger than its mass, suitably normalized by the constants $k$ and $G$.

In what follows, we will see that simple statement has profound implications for particle physics, cosmology, mathematics, and more. But before we get there, let us zoom out a bit and put the Weak Gravity Conjecture into its proper context: the landscape of string theory.

\section{String Theory}\label{sec:ST}

Albert Einstein is famous for his work in two of the most important paradigms in modern physics: quantum mechanics and general relativity. 

Quantum mechanics is what earned Einstein (and many other physicists) a Nobel Prize. It is a theory of very small objects---protons, electrons, atoms, and so on---governed by equations like the Schr\"odinger equation. 

General relativity, on the other hand, is Einstein's theory of gravity. Thus it describes the physics of very heavy objects: galaxies, stars, planets, and the universe taken as a whole. The confirmation of general relativity was what turned Einstein from a successful physicist into a household name.

But what happens if you are faced with objects that are both very small and also very heavy (or, more precisely, very energetic)? For such objects, one must combine quantum mechanics and general relativity into a quantum theory of gravity, or, as it's usually called, a theory of \emph{quantum gravity}. It turns out that this is not so easy to do: the laws of quantum mechanics and the laws of general relativity don't play nicely for sufficiently dense objects, indicating that an entirely new paradigm is needed.

What exactly does ``sufficiently dense'' mean, you might wonder? How often does this actually come up, in practice? The answer is: only in extreme circumstances, which you're unlikely to experience in the real world. Quantum gravity is important for understanding the singularity at the center of a black hole, or the cosmic singularity at the beginning of our universe known as the ``big bang.'' It is important for understanding the period right after the big bang, known as cosmological inflation (see Section \ref{s.implications} for more on this topic). And it is important for understanding fundamental issues facing modern physics, such as the cosmological constant problem.

More broadly, quantum gravity is important because, in a sense, it is the last missing piece of the puzzle of fundamental physics. The first step in the puzzle was the combination of Newtonian mechanics with electromagnetism to give special relativity. Special relativity combines with gravity to give general relativity. Meanwhile, electromagnetism and thermodynamics together give rise to quantum mechanics, which combines with special relativity into relativistic quantum field theory. The last step, then, is to combine quantum field theory with general relativity into a theory of quantum gravity, thereby yielding a complete description of the fundamental laws of nature.

Quantum field theory and general relativity are, by now, among the most well-tested ideas in all of human thought. Quantum gravity, on the other hand, is far more mysterious. However, we aren't completely in the dark when it comes to quantum gravity, and this is where string theory comes in, because...
$$
\boxed{
\textrm{String theory is (by far) the best understood theory of quantum gravity.}}
$$
Indeed, some would go so far as to say that string theory is the \emph{only} mathematically consistent theory of quantum gravity. In this article, we will not take a stand on either side of this debate; for our purposes, the most important thing about string theory is that it is \emph{a} mathematically consistent theory of quantum gravity, which means that we can use it as a toy model to understand what quantum gravity is like.

String theory is called string theory because it is a theory of strings. In particle physics, we are accustomed to thinking of the fundamental constituents---quarks, leptons, gauge bosons, etc.---as being point-like in space. In string theory, however, the fundamental constituents are extended in one dimension, so we call them strings. They have a length, which we call the string length. Strings can be open (with disjoint endpoints) or their endpoints can join together to form a closed string (i.e., a loop).

There are a number of different types of string theory, of which Type I, Type IIA, Type IIB, and heterotic are among the most famous. All of these types of string theory exist naturally in 10 dimensions (1 dimension of time, 9 of space). At first, these different types of string theory appeared to be different theories of quantum gravity, but in the mid 1990s, people realized that these seemingly distinct theories are actually just different limits of a single theory, called M-theory \cite{Witten:1995ex}. M-theory exists naturally in 11 dimensions, and different limits of M-theory reduce to different types of string theory similarly to how one limit of general relativity reduces to Newtonian gravity, while another limit reduces to special relativity. The different types of string theory are thus said to be related by \emph{dualities}, i.e., they represent different descriptions of the same theory. The collection of all of these dualities is called the \emph{string duality web}.

Of course, the world we observe is 4-dimensional (including time), not 10- or 11-dimensional.
Indeed, this is quite clear from the $1/r^2$ falloff of the gravitational force we saw in \eqref{graveq}: this scaling behavior would be replaced by $1/r^{d-2}$ in $d$ dimensions, i.e., $1/r^8$ in 10d and $1/r^9$ in 11d, so the observed gravitational force immediately tells us that we live in 4d.

In string theory, the way to deal with this problem is to curl up, or ``compactify,'' the remaining 6 spatial dimensions into a microscopic, compact manifold, leaving just 4 large dimensions of spacetime. Provided this compact space is small enough, the gravitational effects of the compact dimensions will be negligible, resulting in the correct $1/r^2$ scaling of gravity.

Above, we agreed to remain agnostic about the question of whether or not string theory is the unique theory of quantum gravity. In 10 dimensions, however, the duality web gives us a hint in this direction: what naively seems like a collection of distinct theories is actually a single theory with a collection of different, dual descriptions. Relatedly, string theory in 10 dimensions has no free, dimensionless parameters: the only parameter of string theory is the string length, which has dimensions of length, and there are no other parameters with dimensions of length with which to form a dimensionless ratio. If dimensionless parameters did exist, they could be varied to move from one theory to another. Thus the absence of such parameters indicates the uniqueness of the theory.

This absence of dimensionless parameters in string theory stands in stark contract with the world we observe: the standard models of particle physics and cosmology have many free parameters (e.g. Newton's Gravitational constant, the Planck constant, the cosmological constant, the fine structure constant, the mass of the electron, etc.), which can be used to form many dimensionless ratios. How does one go from string theory in 10d, with no free parameters, to the 4d standard models of particle physics and cosmology that we know and love, which have many? The answer is that, upon compactifying on a 6-manifold, the resulting 4d parameters are related to the details of the compactification, such as the topology, size, and shape of the 6-manifold in question, and the generalized fluxes threading certain submanifolds of this manifold. 

There are millions of eligible 6-manifolds and an exponentially large number of ways to thread fluxes through the submanifolds of one of these manifolds, leading to an exponentially large number of possible solutions to the equations of string theory, and thus an exponentially large number of possibilities for the parameters of the resulting 4d theory. Not every 6-manifold will work, and not every configuration of fluxes will lead to a consistent, stable universe in 4d. We don't know for sure how many solutions there are, but naive estimates include numbers as big as 10$^{500}$ and 10$^{272,000}$ \cite{Taylor:2015xtz}! These estimates should be taken with an enormous grain of salt, but suffice it to say that the evidence points to a vast collection of 4-dimensional solutions to string theory. Each of these solutions represents a different \emph{effective field theory} (EFT), governed by a different choice for the laws of nature in the 4-dimensional universe.\footnote{An effective field theory is a framework for describing a system using quantum field theory. The adjective ``effective'' emphasizes that the field theory is an approximation to a more fundamental theory of quantum gravity. An EFT is reliable at low energies but breaks down at high energies, where it must be replaced by the more fundamental theory.} The set of all EFTs that are consistent with string theory form what is known as the \emph{string landscape}. 

Whether you love the landscape or hate the landscape depends on your perspective. Certainly, though, if your goal is to test string theory experimentally, the landscape is a headache. Simply put, the more possibilities you have, the harder it is to make a unique prediction.

From this perspective, there is some good news: although the string landscape may be very large, it is likely only a small part of an even larger \emph{swampland} of EFTs \cite{Vafa:2005ui, Ooguri:2006in}. EFTs in the swampland have the property that, while they may seem consistent to a low-energy effective field theorist, they are ultimately incompatible with string theory.

\begin{figure}
\centering
\includegraphics[width=100mm, trim={10cm 3cm 10cm 10cm}, clip]{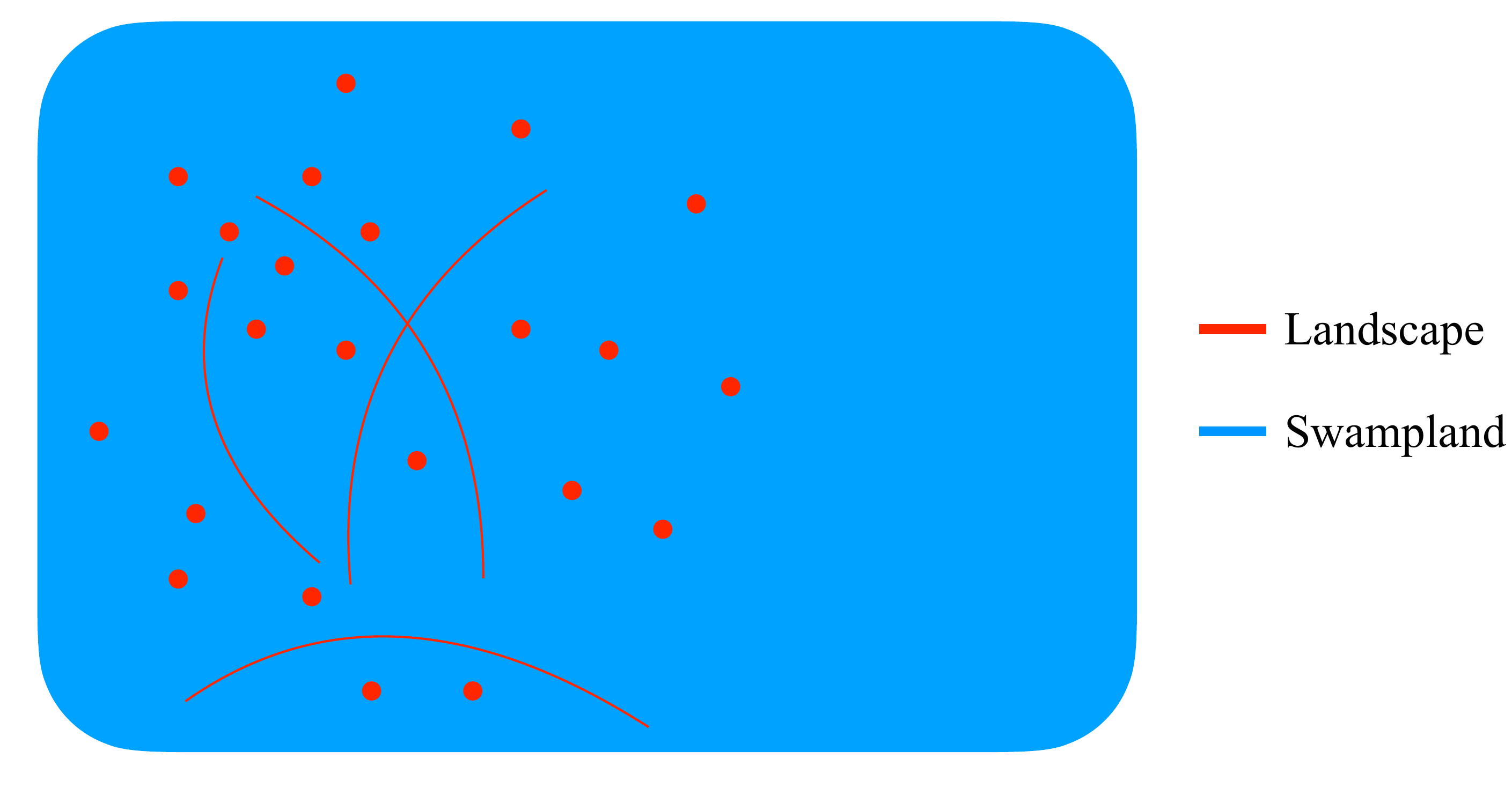}
\caption{Cartoon picture of the landscape and the swampland. The landscape is a chain of some exponentially large number of islands inside a much larger swampland.}
\label{swampmap}
\end{figure}

At present, our ability to map the string landscape and the swampland is in its early stages.  However, to the best of our understanding, the cartoon picture looks a little bit like Figure \ref{swampmap}. The landscape consists of a chain of islands inside a much larger ocean of swampland. In a sense, the Landscape is very large, there are possibly 10$^{500}$ or 10$^{272,000}$ islands in this island chain. On the other hand, there is another sense in which the Landscape is very small. It likely represents only a measure zero subset of the space of all  EFTs.

The goal, then, is to try to delineate the boundary between the landscape and the swampland,  to determine what are the essential, universal features of the EFTs in the string landscape which  distinguish them from the inconsistent EFTs in the swampland. This task is especially important for string theorists who are concerned  with experimental tests of string theory, because if you can show that some model of cosmology or particle physics involves an EFT that lies in the swampland, then experimental verification of that model would rule out string theory. However, even from a purely theoretical perspective, the question of what are the universal features of string theory is a crucial one as we seek to better understand the mysterious paradigm that is quantum gravity. Thus, the task of charting the string landscape is fundamentally a mathematical one, but it often proceeds with an eye towards the real world of particle physics and cosmology.

And this, finally, is where the WGC comes into the story. The WGC is one example of what is known as a “swampland conjecture.” It is a statement which is observed to be true in a vast collection of EFTs in the string landscape, and which is conjectured to be true in \emph{all} EFTs in the landscape.

As with any conjecture in mathematics or physics,  there are three key questions that we must ask of the WGC: 1) What is the precise statement of the conjecture? 2) Is it true? 3) If it is true, what does it imply?

We will now look at each of these questions in turn.

\section{The Weak Gravity Conjecture, Precisely}\label{statement}

In Section \ref{s.invitation}, we introduced the WGC as a statement about the relative size of the repulsive electromagnetic force and the attractive gravitational force between a pair of electrons. This is a useful way to think of the WGC in most cases, but there is a slightly more precise and more general way to define it, with reference to black hole physics. 

Recall that a black hole is a region of spacetime in which gravity is so strong that even light cannot escape. In classical general relativity, a black hole is defined by three numbers: its mass $M$, angular momentum $J$, and charge $Q$. Here and below, we assume that the angular momentum vanishes, $J=0$, leaving us with a charged, \emph{Reissner-Nordstr\"om black hole} of mass $M$ and charge $Q$. Such charged black holes can be sorted into three types:
\begin{enumerate}
    \item Subextremal: $Q< M$.
    \item Extremal: $Q=M$.
    \item Superextremal: $Q> M$.
\end{enumerate}
Here and below, we are working in so-called ``natural units,'' in which various fundamental constants such as Newton's gravitational constant and the speed of light are set equal to 1: $G = c = 4 \pi \varepsilon_0 = 1$.

These three different types of black holes have very different behavior. The case of a subextremal black hole is the most commonplace: The black hole has a time-like singularity hidden behind a light-like event horizon, which is the point of no return. Anything that passes through the event horizon--including light--is trapped inside it. There is a second type of horizon, known as a Cauchy horizon, inside the black hole.

The extremal black hole is similar to the subextremal black hole, except that here the inner (Cauchy) horizon and the outer (event) horizon coalesce into a single horizon.

The superextremal black hole is the most unusual: for $Q > M$, the event horizon disappears entirely, and we are left with a \emph{naked} singularity.

The \emph{(weak) cosmic censorship hypothesis} in general relativity holds that such naked singularities cannot arise from generic initial conditions: any singularity must be hidden from a distant observer by an event horizon. This ensures that no light ray emanating from the singularity can reach a distant observer: the only way to get information about the black hole singularity is to fall past the event horizon.
If this hypothesis is correct, it means that a superextremal black hole cannot form through ordinary dynamical processes; in contrast, a subextremal black hole can certainly be created via the gravitational collapse of a shell of charged matter.

Finally, note that this discussion of charged black holes applies not only to electromagnetism in our universe, but to any gauge force associated with the Lie group U(1). (For more on gauge forces and Lie groups, see the appendix below.) Electromagnetism is the most famous and most relevant example of such a U(1) gauge theory, but a different EFT in the landscape might have a different U(1) gauge group. For instance, some models of dark matter include a ``dark photon,'' which is the gauge boson associated with a different U(1) gauge force, analogous to the ordinary photon for ordinary electromagnetism. The above discussion of subextremal, extremal, and superextremal black holes applies to black holes charged under the dark U(1) just as it does for black holes charged under under electromagnetism.

With this background, we are finally ready to give the precise definition of the WGC:
\vspace{.2cm}
         \begin{namedconjecture}[The Weak Gravity Conjecture]
Given some EFT in the landscape with a U(1) gauge force, there exists a ``superextremal'' particle, i.e., a particle that satisfies\footnote{For the sake of simplicity, we assume without loss of generality that the charge of a given particle or black hole is positive: $q > 0$, $Q>0$. Any particle of negative charge $-q$, such as an electron, has an antiparticle of positive charge $q$, such as the positron.}
\begin{align}
\frac{q}{m} \geq \frac{Q}{M}\big|_{\rm ext}\,,
    \label{WGCeq}
\end{align}
where $Q/M|_{\rm ext}$ is the charge-to-mass ratio of a large, extremal black hole.
            \end{namedconjecture}
    \vspace{.1cm}
\noindent
Let us break down this definition piece by piece. To begin, note that the WGC is a swampland conjecture, which means it deals with EFTs \emph{in the landscape}. It is not hard to write down a Lagrangian for an EFT with a single subextremal particle that violates \eqref{WGCeq}, but the claim of the WGC is that this theory is not consistent with string theory.

Second, note that the term ``superextremal'' in the definition of the WGC is really a shorthand for ``superextremal \emph{or extremal}.'' There are important examples of theories in the string landscape in which the WGC bound \eqref{WGCeq} is exactly saturated, and there are no strictly superextremal particles.

Third, note that the WGC does not require \emph{every} particle in the theory to be superextremal: the claim is merely that there exists at least one. For example, in the case of electromagnetism in our universe, neutrinos are massive ($m \neq 0$) but uncharged ($q = 0$), so they violate the WGC bound \eqref{WGCeq}. However, the WGC is still satisfied because the electron satisfies this bound.

Fourth, in simple cases, the ratio $Q/M|_{\rm ext}$ appearing on the right-hand side of the WGC bound is a constant. By an appropriate choice of units, this constant can be set to 1, and the WGC bound \eqref{WGCeq} is given simply by $q > m$. However, in theories with massless scalar fields, the extremality bound can be modified by changing the values of these scalar fields. In this case, the ratio $Q/M|_{\rm ext}$ is given by some order-one number $\gamma$ in natural units, but $\gamma=\gamma(\phi^i)$ may depend on the massless fields $\phi^i$.

Finally, the definition of the WGC specifies that we are dealing with a ``large'' black hole because the charge-to-mass ratio $Q/M$ of small black holes can be modified slightly relative to the value at infinity. The value in \eqref{WGCeq} should be understood as the limiting value as the size of the black hole goes to infinity, i.e., $\lim_{M \rightarrow \infty} Q/M|_{\rm ext}$.

How does this definition of the WGC compare to the definition given in the Section \ref{s.invitation} in terms of the relative strength of the electromagnetic repulsion and gravitational attraction of a pair of widely separated electrons? In the absence of massless scalar fields, these turn out to be exactly the same:
the condition that a particle is superextremal is precisely the condition that two copies of the particle will repel each other at long distances.
However, this precise correspondence breaks down in the presence of massless scalar fields. Such fields mediate a long-range attractive force,\footnote{A massive scalar field like the Higgs boson also mediates a force between particles, but this force is not a ``long-range'' force because its strength decays exponentially at long distances $r$ as $F \sim \exp(- \alpha m_H r)$, where $m_H$ is the Higgs mass, $\alpha$ is a constant, and we are working in natural units, with $\hbar = c =1$.} and as a result two superextremal particles may attract each other \cite{Palti:2017elp, Heidenreich:2019zkl}; conversely, two subextremal particles may repel each other.

In practice, we expect that massless scalar fields will exist only in supersymmetric theories. In non-supersymmetric theories, quantum corrections will generate a nonzero mass for a particle even if its mass is set to zero classically, but in supersymmetric theories, these quantum corrections cancel between bosons and fermions, and the mass of the scalar field is protected. As a result, for practical applications to the real world (where supersymmetry is broken), the distinction between self-repulsiveness and superextremality is irrelevant. However, most of the evidence for the WGC in string theory comes from supersymmetric EFTs, which feature massless scalar fields; in these theories, the distinction is important to keep in mind.

\subsection{The WGC with multiple photons}\label{ss.multiple}

So far, we have focused on theories with a single U(1) gauge force, with gauge field $A_\mu = (\Phi, A_1, A_2, A_3)$. What happens, though, if the theory has multiple such forces?
In this case, we expect that the WGC should apply to every U(1) in the theory, individually. In a theory with two U(1)s, then, we expect one particle to satisfy the bound \eqref{WGCeq} with respect to the first U(1) and a second particle to satisfy it with respect to the second U(1). However, we can also make a basis change on the U(1) gauge gauge fields:
\begin{align}
   \left(\begin{array}{c}
    (A_\mu^{(1)})' \\
     ( A_\mu^{(2)})'
    \end{array} \right) = \left(\begin{array}{cc}
    \cos \theta & \sin \theta \\
    - \sin \theta & \cos \theta
    \end{array} \right)  \cdot  \left(\begin{array}{c}
      A_\mu^{(1)} \\
      A_\mu^{(2)}
    \end{array} \right) \,.
\end{align}
Denoting the charge of a particle under the $i$th U(1) as $q_i$, this induces a corresponding transformation on the charge of the particle:
\begin{align}
   \left(\begin{array}{c}
   q_1' \\
   q_2'
    \end{array} \right) = \left(\begin{array}{cc}
    \cos \theta & -\sin \theta \\
     \sin \theta & \cos \theta
    \end{array} \right)  \cdot  \left(\begin{array}{c}
    q_1 \\
    q_2
    \end{array} \right) \,,
\end{align}
This basis choice is unphysical: it is merely a relabeling of the degrees of freedom of the theory. As a result, the WGC must be satisfied not only for the original gauge fields $A_\mu^{(1)}$, $A_\mu^{(2)}$, but also for the rotated gauge fields $(A_\mu^{(1)})'$, $(A_\mu^{(2)})'$.

It turns out that there is a simple geometric way to rephrase the above definition of the WGC for a theory with multiple gauge fields \cite{Cheung:2014vva}. To begin, given some particle species, we consider the vector of charges $q_i$ of that particle under each of the U(1) gauge fields in the theory. In a theory with $n$ U(1)s, this vector will have $n$ components, one for each U(1).

\begin{figure}
\centering
\includegraphics[width=110mm, trim={12cm 3cm 12cm 7.5cm}, clip]{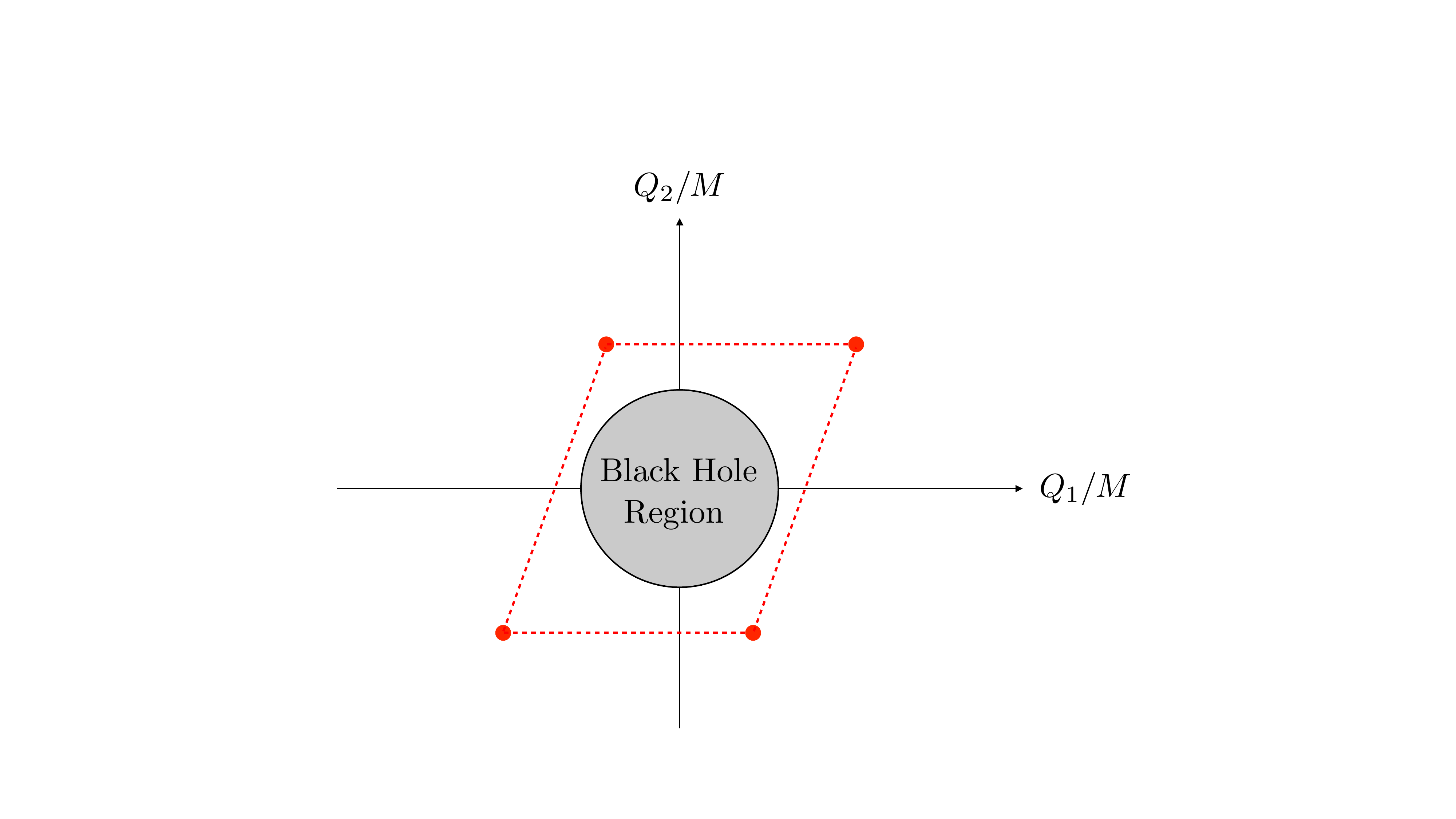}
\caption{The WGC in theories with multiple U(1) gauge fields. The WGC amounts to the statement that the convex hull (red dashed line) of the charge-to-mass vectors of the particles (red dots) should contain the black hole region (gray) in charge-to-mass space. In simple theories, the black hole region is simply the unit ball.}
\label{CHC}
\end{figure}

Next, we define the \emph{charge-to-mass vector} $z_i$ of the particle as
\begin{equation}
 z_i = \frac{q_i}{m}\,.
\end{equation}
With this the WGC is equivalent to the statement that the convex hull of the charge-to-mass vectors of each of the charged particles must contain the black hole region, where subextremal and extremal black holes live. In the simplest case where there are no massless scalar fields, the black hole region is simply the unit ball (in natural units), so the WGC holds that the convex hull of the particle charge-to-mass vectors must contain the unit ball. It is not hard to see that this reduces to the bound \eqref{WGCeq} in the case of a single U(1): The 1-dimensional unit ball is simply an interval of length 2 centered at the origin, and the statement that a 1-dimensional charge-to-mass vector $z = q/m$ lies outside this ball is simply the familiar statement that $q > m$. A 2-dimensional case is shown in Figure \ref{CHC}.

In a theory with massless scalar fields, the black hole region may be modified, taking a different size or shape. However, this black hole region will always contain the unit ball. This means that the WGC can only become stronger by the addition of massless scalar fields to the theory, never weaker.

\subsection{The WGC for strings and branes}

So far, we have focused on the ordinary WGC, which bounds the mass of a particle in terms of its charge. But the WGC admits a natural generalization to higher-dimensional objects charged under higher-dimensional gauge forces.

In electromagnetism, the gauge field $A_\mu = (\Phi, A_1, A_2, A_3)$ transforms under a gauge transformation as
\begin{align}
    A_\mu \rightarrow A_\mu' = A_\mu + \partial_\mu \lambda(x)\,,
\end{align}
where $\lambda$ is a scalar function. Mathematically speaking, $A_\mu$ is an example of a \emph{1-form} gauge field. More generally, we may consider a \emph{$p$-form} gauge field $B_{\mu_1, ..., \mu_p}$, which has $p$ indices. This transforms under a gauge transformation as
\begin{align}
 B_{\mu_1 ,...,\mu_p}  \rightarrow B_{\mu_1 ,...,\mu_p}' = B_{\mu_1 ,...,\mu_p} + \partial_{[\mu_1} \Lambda_{\mu_2 ,...,\mu_p]}(x)\,,
\end{align}
where $\Lambda_{\mu_2,...,\mu_p}$ is a $(p-1)$-form.

The objects charged under an ordinary gauge field are pointlike (i.e., 0-dimensional) particles. In contrast, the objects charged under a $p$-form gauge field are $(p-1)$-dimensional objects, often referred to as \emph{$(p-1)$-branes}. So, a 0-brane is a particle, a 1-brane is a string, and so on. Whereas a particle is characterized by its mass, a higher-dimensional brane is characterized by its \emph{tension}, which has units of mass per unit length.

The ordinary WGC places a lower bound on the charge-to-mass ratio of a particle charged under a 1-form gauge field. This admits a natural generalization to the case of a $p$-form gauge field, requiring a $(p-1)$-brane of tension $T$, charge $q$, which satisfies
\begin{align}
\frac{q}{T} \geq \frac{Q}{T} \Big|_{\rm ext} \sim 1\,,
\label{WGCpform}
\end{align}
where now $Q/T |_{\rm ext}$ represents the charge-to-tension ratio of an extremal \emph{black brane} charged under the $p$-form of interest \cite{Arkanihamed:2006dz}.

This definition of the $(p-1)$-brane WGC generally makes sense for $p = 1, ..., d-3$, where $d$ is the dimension of spacetime. However, it may be possible to extend the definition to the $p=0$ case as well. Here, the $0$-form gauge field in question is a periodic scalar field $\phi$, i.e., a scalar field $\phi$ subject to the identification $\phi \equiv \phi + 2 \pi f$ for some constant $f$. The charged objects are $(-1)$-branes, which may seem strange, but can be understood as follows: recall that a particle is 0-dimensional, which means that it is localized in space, but still it travels through time, i.e., it has a 1-dimensional worldline. In contrast, a $(-1)$-brane is localized in both space \emph{and} time. It is known as an \emph{instanton}, because it is associated with just a single instant in time. In place of a mass or tension, an instanton has a quantity known as an action, $S$, which is dimensionless.

The instanton version of the \eqref{WGCpform} is a bit nebulous, because there is no such thing as a ``black instanton,'' so $\frac{Q}{S} \big|_{\rm ext}$ is not a well-defined quantity. The typical response to this is simply to allow the instanton version of the WGC (also referred to as the axion WGC) to be a nebulous statement, which requires 
\begin{align}
\frac{q}{ S} \geq c\,,
\label{axionWGCeq}
\end{align}
where $c \sim O(1)$ is some order-one number in natural units, and the minimum charge of the instanton is given by the inverse of the periodicity parameter, $q = 1/f$. The question of what the precise value of $c$ should be in this expression is still open \cite{Heidenreich:2015nta, Hebecker:2016dsw, Hebecker:2018ofv, Andriolo:2020lul}. Answering this question is important because, as we will see below, the instanton version of the WGC is arguably the most important version for cosmological purposes.

\section{Evidence for the Weak Gravity Conjecture}

The WGC is a conjecture, not a theorem; it has yet to be proven in full generality. However, several lines of evidence point to its validity. In this section, we will discuss a few of them.

\subsection{Heuristic motivation: black hole evaporation}\label{ss.H}

Let us begin with the original argument given in favor of the WGC \cite{Arkanihamed:2006dz}. We will see that this argument has a big hole; it is better understood as a heuristic argument than a serious attempt at proof. Nonetheless, this argument will help us think about the WGC from a more physical perspective, and ultimately it may point us in the right direction to prove the WGC.

\begin{figure}
\centering
\includegraphics[width=120mm, trim={5cm 0cm 5cm 0cm}, clip]{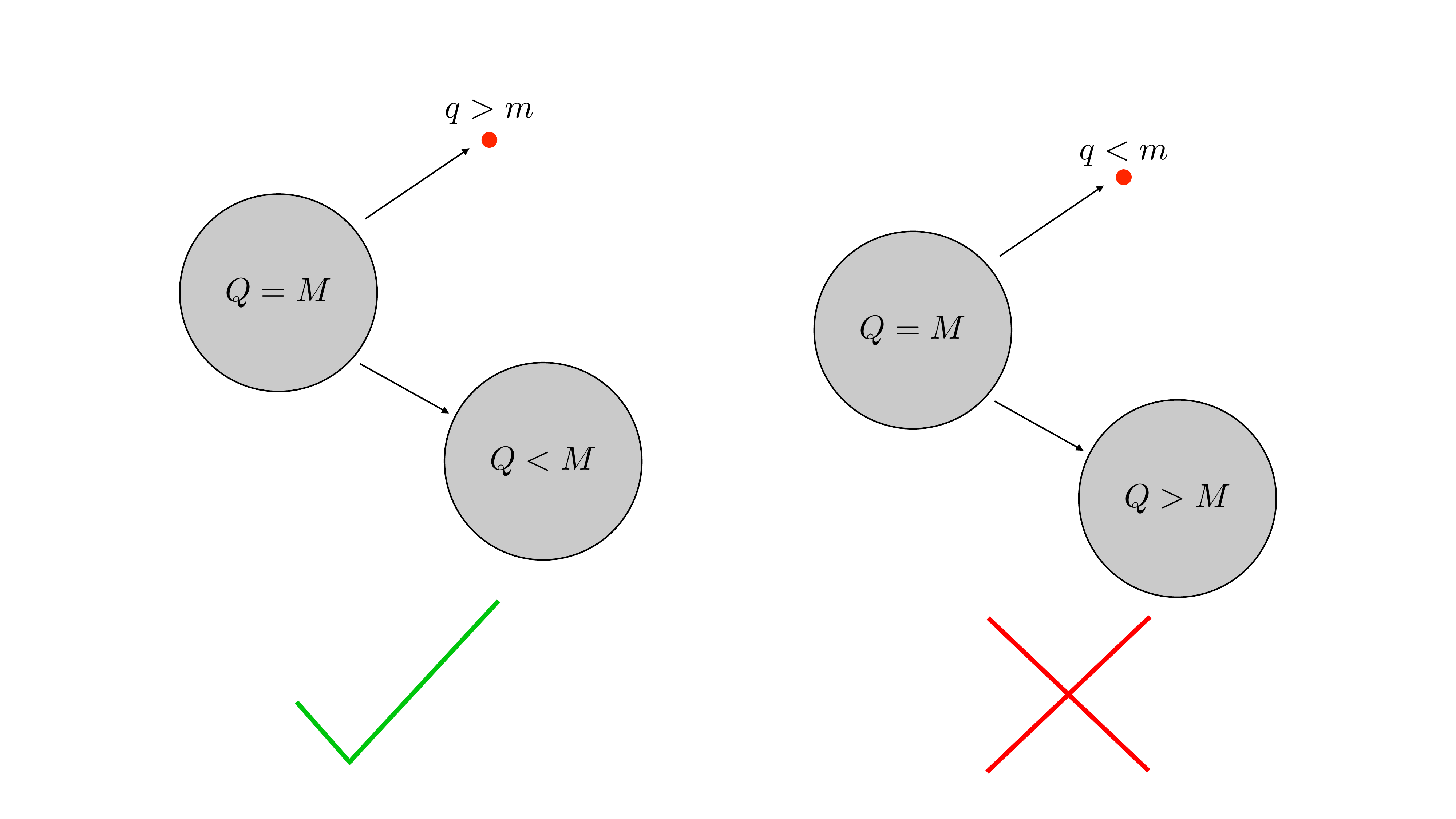}
\caption{Extremal black hole evaporation. In a theory that satisfies the WGC, an extremal black hole can decay by emitting subextremal particles. In a theory that violates the WGC, however, an extremal black hole cannot decay without violating the extremality bound, and thus violating cosmic censorship; instead, we expect that such a black hole will be stable.}
\label{BHevap}
\end{figure}

The argument runs as follows: consider a large, extremal Reissner-Nordstr\"om black hole of mass $M$ and charge $Q=M$. If the WGC is satisfied, then there exists a particle of mass $m$ and charge $q \geq m$, and the black hole can decay through a process known as \emph{Hawking evaporation} by emitting one of these particles,\footnote{More precisely, this evaporation process should be called \emph{Schwinger pair production}, as it involves the emission of a charged particle.} as shown in Figure \ref{BHevap} (left). The resulting black hole will then have mass $M' \approx M - m$ and charge $Q' = Q - q$ and will be subextremal, $M' \geq Q'$. This decay process can then repeat itself ad naseum, until the black hole has evaporated completely. Indeed, this process is roughly what would happen to a hypothetical extremal black hole in our own universe: it would emit electrons until it had shed all of its charge, and eventually it would decay to zero size by emitting uncharged Hawking radiation.

But what if the WGC is violated? Then, as shown in Figure \ref{BHevap} (right), the extremal black hole in question could only shed its charge by emitting a subextremal particle, with $m > q$. After this, the resulting black hole would be superextremal, with $M' = M - m < Q' = M- q$. As discussed above, this superextremal black hole would introduce a naked singularity to the spacetime, and thus this decay process would violate the cosmic censorship hypothesis.

Thus, if we assume that there are no superextremal black holes in the theory, the extremal black hole is kinematically stable: any decay process is forbidden, and the extremal black hole just sits there indefinitely. Thus, if stable, extremal black holes are forbidden, we conclude that the WGC must be satisfied.

The problem with this argument is that no one has yet come up with a fully convincing reason as to why stable extremal black holes present a problem for quantum gravity. Indeed, there exist supersymmetric theories in which the WGC is saturated, i.e., $q=m$: in these cases, an extremal black hole can decay only at threshold, and the lifetime of such a decay process is infinite: extremal black holes are marginally stable. To finish off the argument and prove the WGC, then, we would need to come up with an argument against stable, extremal, non-supersymmetric black holes that leaves open the possibility of stable, extremal black holes.

\subsection{Top-down evidence: Examples in string theory}\label{ss.TOP}

At this point, the strongest evidence for the WGC arguably comes from many examples in string theory, and the lack of a counterexample. In this subsection, we detail one illustrative example in which the WGC is satisfied: heterotic string theory.

There are actually two different types of heterotic string theory, which are distinguished by their gauge groups: one has SO(32) gauge group, while the other has E$_8 \times $ E$_8$. Everything we have to say about heterotic string theory will apply to both types, but for concreteness we will focus on the SO(32) case. Further details on these groups can be found in the appendix.

Both types of heterotic string theory exist naturally in ten dimensions,
but as usual, we may compactify six of these dimensions on a 6-manifold to get a 4d EFT. For simplicity, let us focus on the simplest 6-manifold: a 6-dimensional torus $T^6$, which may be thought of as the direct product of six circles, generalizing the construction of the ``doughnut'' $T^2$ as a product of two circles.

In the process of compactifying to 4d, we may break the SO(32) gauge group down to its maximal abelian subgroup, U(1)$^{16}$. This is a collection of 16 U(1)'s, i.e., 16 different types of electromagnetism, each with their own photon.

\begin{figure}
\centering
\includegraphics[width=120mm, trim={10cm 1cm 10cm 11cm}, clip]{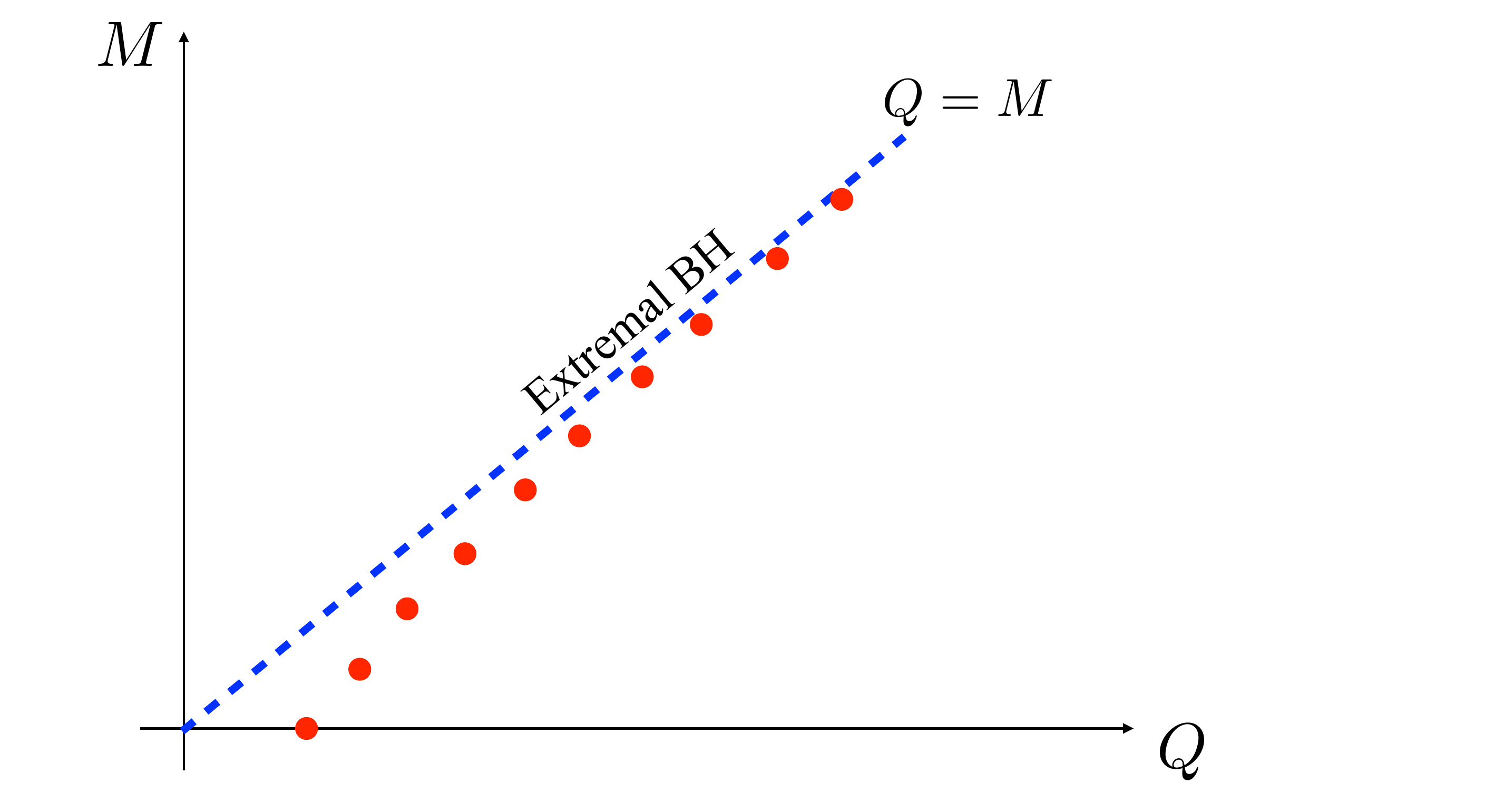}
\caption{Spectrum of charged particles in SO(32) heterotic string theory. The lightest charged state is massless, whereas particles of large charge asymptote to the black hole extremality bound.}
\label{heterotic}
\end{figure}

For concreteness, let us focus here on one particular U(1) gauge force (any other U(1) will be more or less equivalent). For any positive integer $n$, the theory has a particle of charge $q = g n$ and mass
\begin{equation}
    m^2 = g^2 (n^2 - 1) \,,
    \label{heteq}
\end{equation}
where $g$ is the coupling constant of the U(1) gauge force (e.g., $g \approx 0.3$ for electromagnetism in our universe).
This spectrum of particles charged under this U(1) is shown in Figure \ref{heterotic}. The lightest charged particle is massless, so clearly it satisfies the WGC bound, as $q/m \rightarrow \infty$. However, as you can see from the figure, there is actually a whole \emph{tower} of superextremal particles of increasing charge and increasing mass, which asymptote to the extremality bound $Q=M$. Indeed, for very large charge, the states in this tower represent the extremal black holes themselves, which have $M \gg 1$ in natural units.

This example is illustrative of a far more general phenomenon in string theory: typically, the WGC is satisfied not only by a single particle, or even by a finite number of particles, but rather by an infinite tower of superextremal objects of increasing charge and increasing mass. At small charge, these objects can be thought of as particles, like the electron for electromagnetism. At very large charge, they can be thought of as extremal black holes, whose charge-mass-ratio is greater than or equal to the charge-to-mass ratio of an infinitely large extremal black hole, i.e.,
\begin{align}
    \frac{Q}{M}\big|_{\textrm{ext}}^{\textrm{finite M}}  \geq 
 \lim_{M \rightarrow \infty} \frac{Q}{M}\big|_{\rm ext} \,.
\end{align}
This phenomenon is sometimes elevated to a conjecture, called the \emph{tower Weak Gravity Conjecture} \cite{Heidenreich:2015nta, Andriolo:2018lvp}, which holds that every U(1) must have such a tower of superextremal states. Clearly, this conjecture implies the ordinary WGC.

Here, we have focused on one simple example, where $SO(32)$ heterotic string theory is compactified on $T^6$. However, the WGC (and the stronger tower WGC) have been verified in a wide array of string compactifications on more complicated 6-manifolds \cite{Heidenreich:2016aqi, Lee:2018urn,Klaewer:2020lfg,Lee:2019tst, Cota:2020zse, Alim:2021vhs, Gendler:2022ztv}. Notably, \cite{Heidenreich:2016aqi, Heidenreich:2024dmr} proved that the tower WGC holds for any U(1) gauge group that arises from perturbative physics in string theory, leaving just the non-perturbative cases yet to prove in full generality.

\subsection{Bottom-up evidence: EFT arguments}

The vast array of string compactifications offer compelling evidence for the WGC, but ultimately it is difficult to imagine a proof of the WGC coming from such means: at present, we have a relatively good handle on certain classes of supersymmetric EFTs in the string landscape; we know very little about any non-supersymmetric EFTs in the landscape, let alone $10^{500}$ or $10^{272,000}$ of them!

As a result, if we are going to prove the WGC, it is likely going to come from a ``bottom-up'' perspective, as some sort of low-energy consistency condition on black hole physics. While no one has yet come up with a completely convincing proof, there seems to be a general consensus on the form such a proof would take.

\begin{figure}
\centering
\begin{subfigure}{.45\linewidth}
\includegraphics[width=80mm, trim={13cm 1cm 13cm 11cm}, clip]{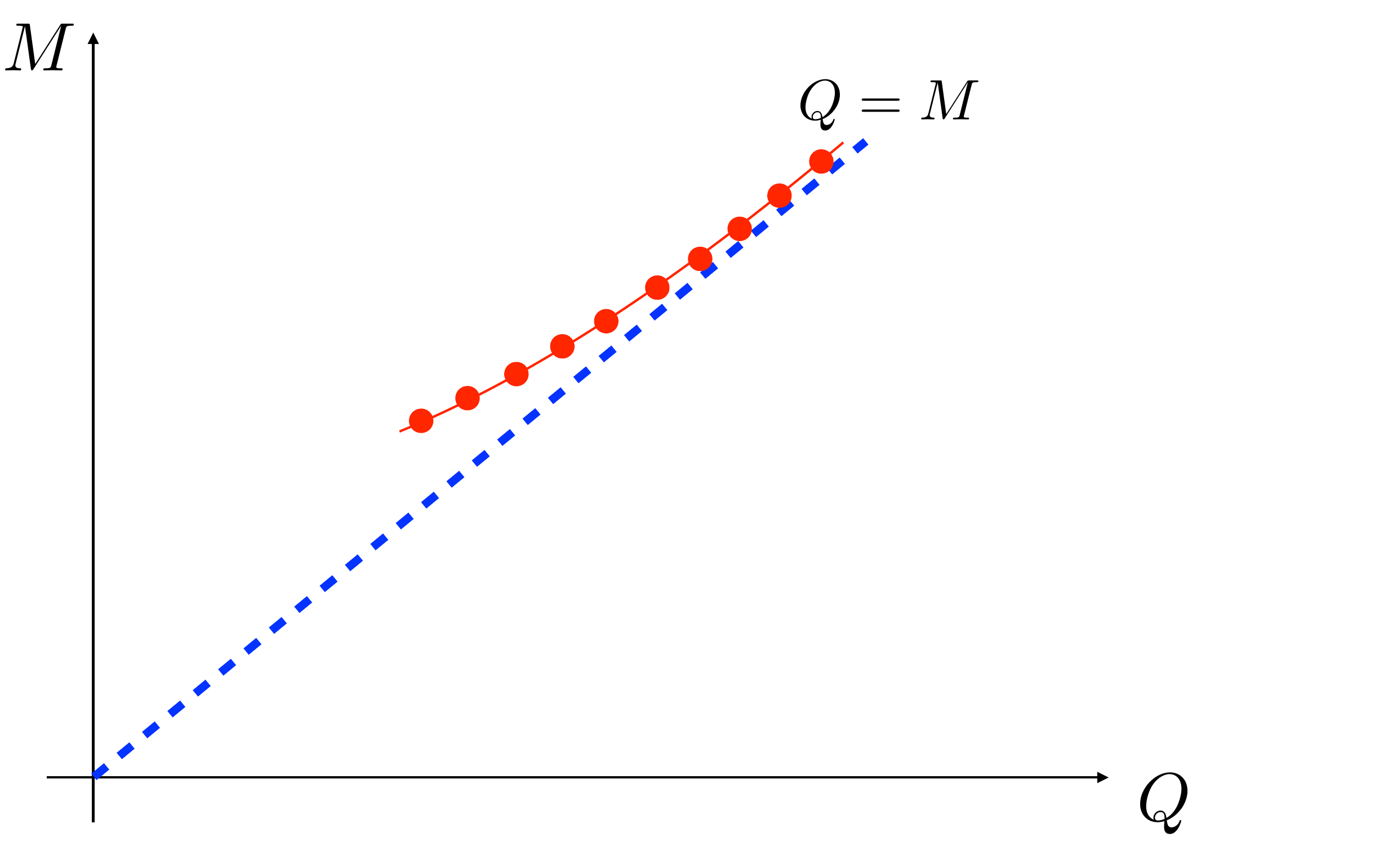}
\caption{$\varepsilon(M) < 0$}
\label{badBH}
\end{subfigure}
\begin{subfigure}{.45\linewidth}
\includegraphics[width=80mm, trim={13cm 1cm 13cm 11cm}, clip]{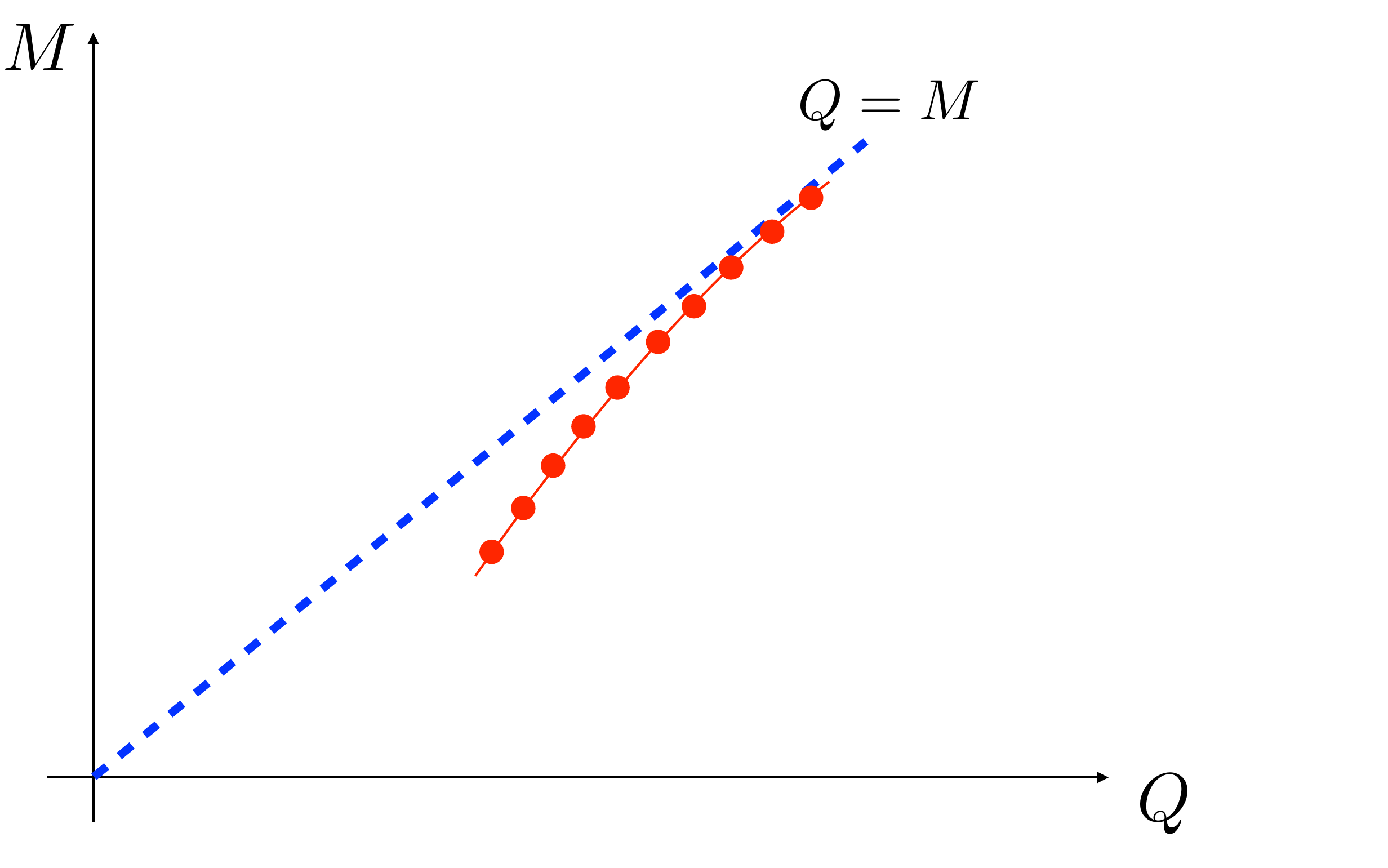}
\caption{$\varepsilon(M) > 0$}
\label{goodBH}
\end{subfigure}
\caption{Spectrum of extremal black holes for $\varepsilon(M) <0$ (left) and $\varepsilon(M) > 0$ (right). Known examples in the landscape have $\varepsilon(M) \geq 0$ for large $M$, which favors the tower version of the WGC.}
\label{goodbadBH}
\end{figure}

Recall from our discussion above that the charge-to-mass ratio of a finite-sized black hole can, in general, depend on the size of the black hole \cite{Kats:2006xp}:
\begin{equation}
    \frac{Q}{M}\Big|_{\rm ext}^{\textrm{finite M}} = 1 - \varepsilon(M)\,,
    \label{epeq}
\end{equation}
where $|\varepsilon(M)| \ll 1$ is a function of the mass $M$ of the extremal black hole satisfying $\lim_{M \rightarrow \infty} \varepsilon(M) = 0$. If $\varepsilon(M)$ is negative, then an extremal black hole does not satisfy the WGC bound \eqref{WGCeq}. In order to satisfy the WGC in this case, we would need additional, superextremal particles of small charge. This possibility is shown in Figure \ref{badBH}.

However, if $\varepsilon(M)$ is positive (Figure \ref{goodBH}), then an extremal black hole of mass $M$ satisfies the bound \eqref{WGCeq}, and the WGC is satisfied! At least, the letter of the law is satisfied, though perhaps not the spirit of it: after all, the original motivation of the WGC was to ensure that black holes can decay by emitting superextremal particles, so it isn't very satisfying if the only superextremal states in the theory are themselves black holes.

Nonetheless, a proof of $\varepsilon(M) \leq 0$ for some $M$ would suffice to prove the WGC in its mildest form, and as a result many works have sought to prove that consistency of black hole thermodynamics/EFT requires 
$\varepsilon(M) \geq 0$ \cite{Shiu:2016weq, Shiu:2017lX, Fisher:2017dbc, Cheung:2018cwt}. A proof of $\varepsilon \geq 0$ would further lend support to the tower WGC introduced in Section \ref{ss.TOP}, which requires the existence of both superextremal particles at small charge as well as extremal black holes with $\varepsilon(M) \geq 0$ at large charge. Indeed, comparing \eqref{epeq} with \eqref{heteq} for $n \gg 1$ (or Figure \ref{goodBH} with Figure \ref{heterotic}), we see that the spectrum of black holes in heterotic string theory on $T^6$ has (in the lowest order approximation) the correct sign $\varepsilon > 0$. Thus, a general proof of $\varepsilon \geq 0$ would go a long way towards confirming the tower WGC, and conversely an example of a theory in the landscape with $\varepsilon(M) < 0$ would be a big surprise, forcing us to rethink our understanding of the WGC and the string landscape.

But let us suppose that we are interested in the spirit of the WGC, not merely the letter of the law. Is there a reason to believe that the WGC should be satisfied by a light particle, not merely a finite-sized black hole?

Several lines of reasoning suggest the answer to this question is yes; for lack of space, we will focus on one particular one: the connection between the WGC and the cosmic censorship hypothesis (defined above in Section \ref{statement}).

Already, the heuristic argument of Section \ref{ss.H} suggests that such a connection should exist. However, the arguments of \cite{Crisford:2017gsb,Horowitz:2019eum} make this connection far more precise. In \cite{Horowitz:2016ezu, Crisford:2017zpi}, Horowitz, Santos, Way, and Crisford found examples of spacetimes that violate the cosmic censorship hypothesis in theories with U(1) gauge fields. However, following up on a suggestion of Vafa, Crisford, Horowitz, and Santos subsequently showed that these violations disappear upon adding a superextremal scalar field to theory \cite{Crisford:2017gsb}. In other words, cosmic censorship is restored precisely when the WGC is satisfied!

In subsequent work \cite{Horowitz:2019eum}, Horowitz and Santos showed that this precise connection between the WGC and the cosmic censorship hypothesis persists in theories with massless scalar fields (where the WGC bound \eqref{WGCeq} is modified by the massless scalars) and in theories with multiple U(1) gauge fields (where the WGC becomes equivalent to the convex hull condition of Section \ref{ss.multiple}). This remarkable connection has a simple explanation: if we assume that the cosmic censorship hypothesis is true for EFTs in the landscape, then any EFT that violates the hypothesis must lie in the swampland. In the scenario of \cite{Crisford:2017gsb,Horowitz:2019eum}, EFTs that violate the WGC also violate cosmic censorship. So, if cosmic censorship is true in the landscape, then the WGC must be true as well.

Of course, this argument for the WGC still requires us to prove that the cosmic censorship hypothesis is true in the landscape, which could be viewed as a bit of a lateral move: we have exchanged one conjecture for another. Nonetheless, in light of old arguments for cosmic censorship from numerical studies of black hole mergers and recent arguments for cosmic censorship in wide swaths of the landscape \cite{Engelhardt:2019btp}, this censorship connection seems like an important piece of evidence in favor of the WGC.

\section{Implications of the Weak Gravity Conjecture}\label{s.implications}

The implications of the WGC spread out in many different directions. In the case of a string compactification, the charge and mass of a charged particle may be related to the size and shape of the compactification manifold; as a result, the WGC translates to a geometric statement about 6-manifolds, which has been verified in many examples.

Using the \emph{AdS/CFT correspondence}, which relates a quantum gravity theory in $d+1$ dimensions to a non-gravitational quantum theory in $d$ dimensions, the WGC translates into a non-gravitational statement, which similarly can be verified in examples.

Ultimately, though, we would like not only to understand string theory at a formal, mathematical level, but also to connect it to observable physics. This is the primary driving force behind the recent wave of excitement over the WGC: its implications for particle physics and cosmology.

A number of such implications have been discussed in the literature, of varying degrees of rigor, precision, and significance. In this section, I will focus on one of the first and (in my humble, biased opinion) most significant applications of the WGC to cosmology: axion models of cosmological inflation.

\begin{figure}
\centering
\includegraphics[width=110mm, trim={12cm 3cm 12cm 10cm}, clip]{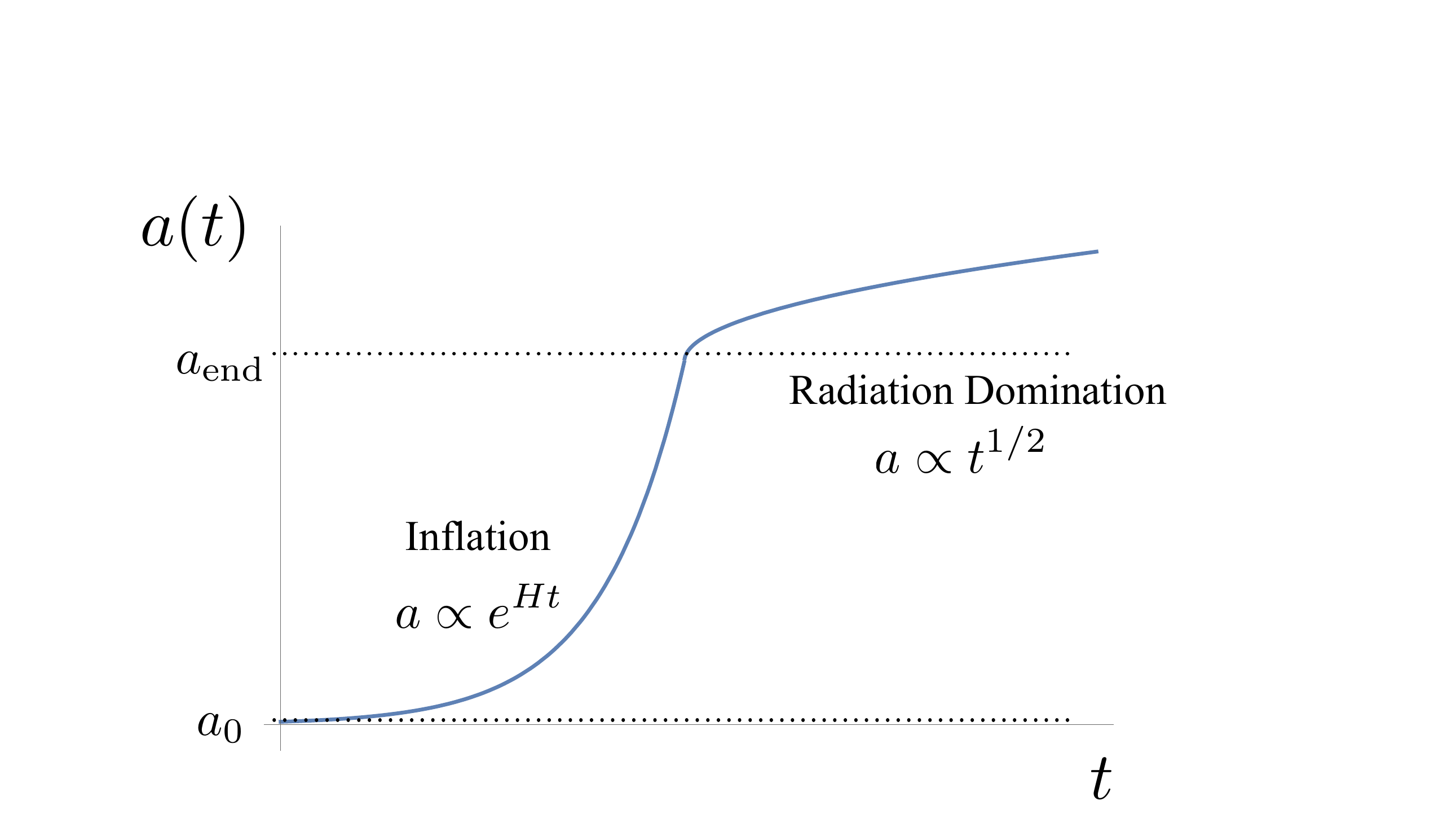}
\caption{The early expansion history of the universe. Inflation posits a period of exponential growth of the universe, as measured by the \emph{scale factor} $a(t)$. Eventually, this exponential growth stops, and the universe continues to grow at a much slower rate.}
\label{inflationplot}
\end{figure}

Cosmological inflation, or simply \emph{inflation}, is a postulated period of exponential growth in the early universe. According to inflation, the universe in its very earliest moments was undergoing a period of explosive, exponential growth, i.e., inflating like a balloon. In a tiny fraction of a second, inflation occurred, then stopped, and the universe continued to expand but at a much slower rate (see Figure \ref{inflationplot}).

Microscopically, inflation can be envisioned as a ball rolling down a hill with friction. In this analogy, the role of the ball is played by a scalar field $\phi$ known as an inflaton, which to good approximation is homogenous in space and depends only on time, $\phi = \phi(t)$. The hill is an inflationary potential $V(\phi) > 0$, and friction is an effect caused by the expansion of the universe, known as Hubble friction. The equation of motion governing the evolution of the inflaton is given by
\begin{equation}
\ddot \phi + 3 H \dot  \phi + \partial_\phi V  = 0\,,
\end{equation}
where one dot denotes a derivative with respect to time, two dots denotes two derivatives with respect to time, $\partial_\phi V$ is the derivative of $V$ with respect to $\phi$, and 
\begin{equation}
H^2 =  \frac{\rho}{3} = \frac{1}{3} \left(  \frac{1}{2} \dot \phi^2 + V(\phi) \right)\,..
\end{equation}
Here, $\rho$ is the energy density, which is split into a kinetic term $\dot \phi^2/2$ and a potential term $V$, similar to a ball rolling down a hill.

Inflation---exponential growth of the universe---occurs when the \emph{Hubble parameter} $H$ is approximately constant in time; this requires the potential $V(\phi)$ to dominate over the kinetic term $\dot \phi^2/2$, which in turn implies an upper bound on the first and second derivative of the potential
\begin{equation}
\frac{|\partial_\phi V|}{V} \ll 1\,,~~~ - \frac{\partial_\phi \partial_\phi V}{V} \ll 1\,.
\end{equation}
So, successful inflation requires a potential which satisfies these conditions for a suitably large range of the field space as the inflaton rolls down its hill; eventually, the conditions are violated, the inflaton starts rolling quickly, and inflation ends: the universe continues to expand, but not at an exponential rate. Any potential which takes this form--typically involving some sort of a plateau followed by some sort of cliff--represents a distinct model of inflation.

\begin{figure}
\centering
\includegraphics[width=80mm, trim={20cm 3cm 20cm 20cm}, clip]{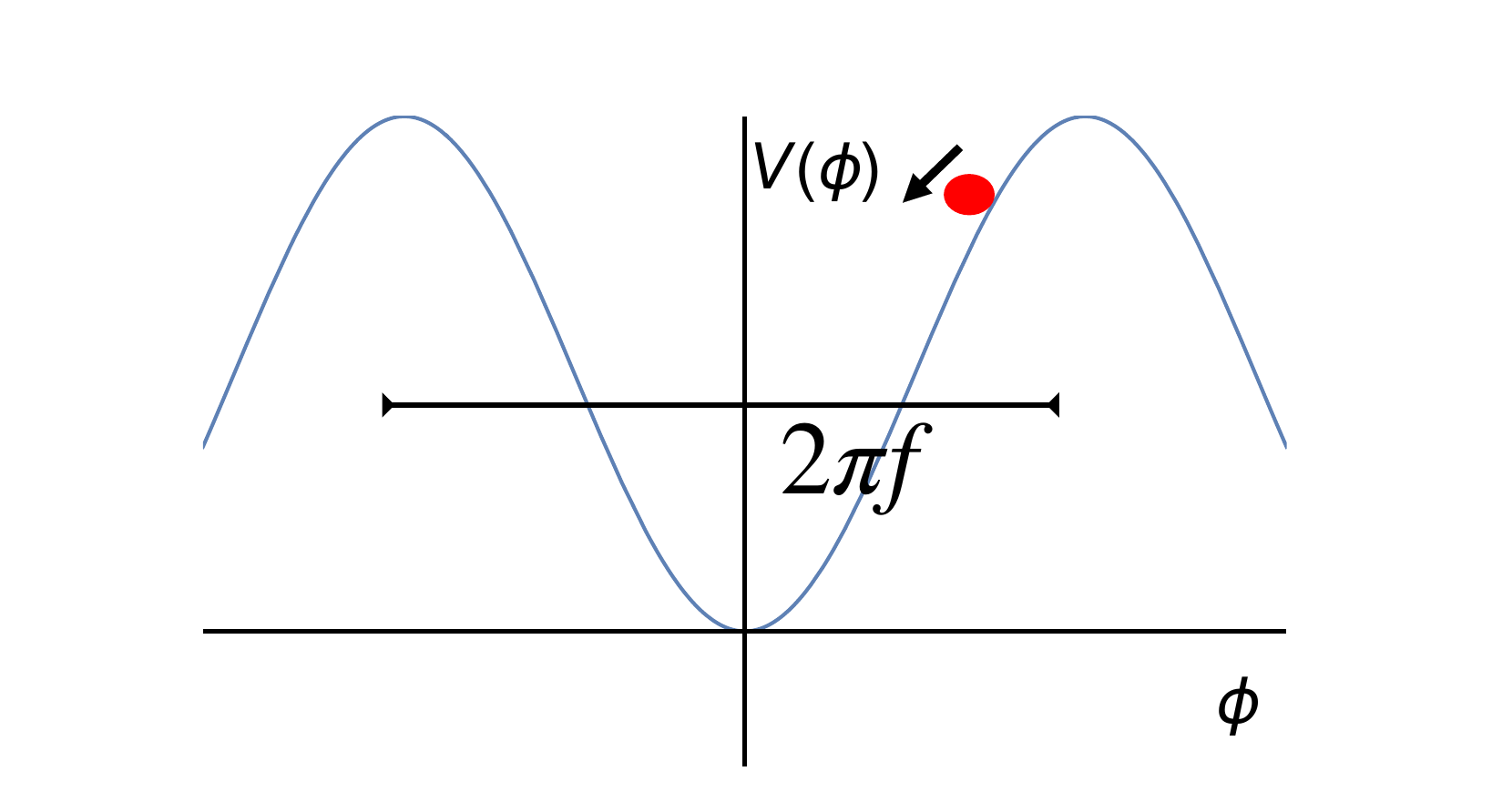}
\caption{Inflationary potential for natural inflation. Inflation occurs while the inflaton field rolls slowly along the top of the hill, then ends as it picks up speed near the minimum. Successful natural inflation requires the cosine well to be large, $f \gtrsim 10$, whereas the WGC requires $f \lesssim 1$.}
\label{naturalinflation}
\end{figure}

At a broad, paradigmatic level, the experimental evidence for inflation is substantial. However, no individual model of inflation is problem-free; such models usually require a fine-tuning of the initial conditions of the inflaton field, a fine-tuning of the potential $V(\phi)$, or both. However, one popular model of inflation---which attempts to get around both of these issues---is known as \emph{natural inflation} \cite{Freese:1990rb}. This model involves a periodic scalar field, also known as an axion, with periodicity $\phi \equiv \phi + 2 \pi f$. Its potential is generated by instantons and takes the sinusoidal form
\begin{equation}
V(\phi) = V_0 e^{-S} \left(  1 - \cos (\frac{\phi}{f}) \right) + O(e^{-2S})\,,
\end{equation}
where the $O(e^{-2S})$ terms come from the next harmonic $\sin(2 \phi)$. Here, inflation occurs when the inflaton rolls slowly near the peak of the cosine term at $\phi \approx \pi f + 2 \pi n f$, and it ends once the field starts rolling quickly, eventually settling into a minimum at $\phi \approx 2 \pi n f$, as shown in Figure \ref{naturalinflation}.

To agree with experimental observations, natural inflation requires $f \gtrsim 10$. Meanwhile, exponential suppression of the higher harmonics requires $S \gg 1$, which together implies $f S \gg 1$.
 But, as we saw above in \eqref{axionWGCeq}, this violates the instanton version of the WGC: natural inflation is in tension with the WGC!
 
 In principle, there are ways to get around this bound, using more complicated models of natural inflation than the simple one considered here, which both satisfy the WGC as well as the requirements on $f$ and $S$. But then, the question becomes: does string theory allow these more complicated models? Or might we satisfy the WGC at the expense of violating some other consistency condition on the string landscape?
 
So far, the evidence suggests the latter: all attempts to produce an axion-instanton system with $f S > 1$ in string theory have failed \cite{Banks:2003sx, Rudelius:2014wla, rudelius:2015xta, Montero:2015ofa, Bachlechner:2015qja, Junghans:2015hba, Palti:2015xra, Conlon:2016aea, long:2016jvd}. Even invoking multiple axions \cite{Liddle:1998jc, Kim:2004rp, dimopoulos:2005ac} doesn't seem to work, as it violates the convex hull condition discussed in Section \ref{ss.multiple} \cite{Rudelius:2014wla, rudelius:2015xta, Montero:2015ofa, Brown:2015iha,Brown:2015lia, Heidenreich:2015wga}. If natural inflation is consistent with string theory, it has yet to be found.

\section{Conclusions}

In this brief introduction to the WGC, we have covered a lot of ground. We have explained the necessity of quantum gravity and the importance of string theory, we have introduced the string landscape and the swampland, and we have explored how the WGC fits into this picture as a candidate consistency criterion for distinguishing the landscape from the swampland. From here, we examined the definition of the WGC in detail, including its relation to black hole physics and its generalization to theories with higher-form gauge fields and multiple gauge fields.

We next looked briefly at the vast body of literature on evidence for the WGC and implications of the WGC. We highlighted evidence for the WGC from string theory, an intriguing connection between the WGC and the cosmic censorship hypothesis, and the importance of the WGC for natural inflation, among other things.

It is remarkable that the simple observation that a pair of electrons will repel (rather than attract) each other could touch on so many different topics at the cutting edge of modern high-energy physics, and in this article we have really only seen the tip of the iceberg. Given further space, we could explore more evidence for the WGC, the attempted proofs of the WGC, and its connections with other areas of particle physics, cosmology, mathematics, and more. In this article, we touched on the tower WGC, but there are other proposed variants of the WGC that are plausibly true and potentially important \cite{Heidenreich:2015nta, Heidenreich:2016aqi, Ooguri:2016pdq}. For a more thorough review of the WGC, see \cite{Harlow:2022ich}.

And yet, even this apparent iceberg turns out not to be an iceberg at all, but rather the tip of an even larger iceberg called the \emph{swampland program} \cite{Vafa:2005ui, Brennan:2017rbf, Palti:2019pca}, which aims to identify universal features of EFTs in the landscape and delineate it from the swampland. The WGC is one candidate for such a universal feature, but there are many others, each of which could easily fill a \emph{Contemporary Physics} article in its own right. As a whole, the swampland program has yet to achieve its ultimate goal of bridging the gap between string theory and experiment, but it has already changed the way that string theory is studied, and most likely the best is yet to come.

\section*{Acknowledgements}

We are grateful to Muldrow Etheredge, Matthew Reece, and Robert Rudelius for their comments on a draft. The work of TR was supported in part by STFC through grant ST/T000708/1.

\appendix

\section{Gauge Theory and Lie Groups}

An effective field theory, or EFT, is a mathematical description of a physical system. The dynamics of such a system can often be described in terms of a \emph{Lagrangian} $\mathcal{L}$, which is a mathematical function of the degrees of freedom of the system, represented by different \emph{fields}. The simplest example of such a field is a \emph{scalar field}, often denoted by the Greek letter $\phi$, which assigns a number to every point in spacetime. A familiar example of a scalar field is temperature: to a given place on earth $\mathcal{x}$ at a given time $t$, we may assign a temperature $T=T(t, \mathcal{x})$. A more fundamental example of a scalar field is the celebrated Higgs boson.\footnote{Indeed, all fundamental scalar fields are bosons rather than fermions.}

Symmetries play a key role in the study of EFTs. Some symmetries are discrete, like the reflection symmetry that flips the two sides of an isosceles triangle. Other symmetries are continuous, such as the rotations that act on a circle. The set of symmetry transformations forms a mathematical object known as a \emph{group}. In EFTs, symmetries are realized by transformations of the fields that keep the Lagrangian invariant. For example, the Lagrangian of a non-interacting, complex-valued scalar field $\varphi$ of mass $m$ in four dimensions takes the form
\begin{equation}
\mathcal{L} =  \frac12 \left(    \dot \varphi^\dagger   \dot \varphi  -  (\vec \nabla \varphi^\dagger) \cdot (\vec \nabla \varphi)     - m^2 \  \right)\,,
\end{equation}
where $\vec \nabla$ is the gradient, dot denotes a derivative with respect to time, and $\varphi^\dagger$ denotes the complex conjugate of $\varphi$. One can check that this Lagrangian is invariant under the symmetry transformation
\begin{align}
\varphi \rightarrow e^{i \alpha} \varphi\,,~~~\alpha \in [0, 2\pi)\,.
\end{align}
I.e., $\mathcal{L} \rightarrow \mathcal{L}$ under this transformation of the field $\varphi$. Here, the parameter $\alpha$ is circle-valued, which tells us that the symmetry group is the Lie group U(1). The fact that $\varphi$ transforms nontrivially tells us that it is \emph{charged} under the symmetry.

A \emph{gauge symmetry} is a special type of symmetry in an EFT. One important consequence of a gauge symmetry is the existence of a gauge field, which mediates a force between two charged particles. In the case of electromagnetism, the gauge field is the usual electromagnetic 4-potential $A_\mu = (\Phi, A_1, A_2, A_3)$. The particle associated with this gauge field is the photon, which mediates a force (the electromagnetic force) between two charged particles, e.g. two electrons.

Thus, in EFT, gauge forces are associated with symmetry groups. The symmetry group for electromagnetism is U(1)---the group of rotations of a circle. U(1) is one example of a \emph{Lie group}, which is a group that is also a smooth manifold. Another example of a Lie group is SO(3)---the group of rotations of a 2-sphere. Yet another is SU($N$), the set of $N \times N$ unitary matrices of determinant 1. Given any such Lie group $G$, we can construct a \emph{gauge theory}, i.e., an EFT with gauge group $G$. For example, the standard model of particle physics has a gauge symmetry group given by the direct product of three Lie groups: SU(3)$\times$SU(2)$\times$U(1): the SU(3) factor is associated with a gauge force called the \emph{strong force}, the SU(2) is associated with a gauge force called the \emph{weak force}, and the U(1) is associated with electromagnetism.

Some Lie groups come in infinite families; for instance, SO($N$), $N \geq 2$ represents the set of $N \times N$ orthogonal matrices of determinant 1. There are some Lie groups, however, which do not fall into such infinite families---examples include the exceptional Lie groups E$_6$, E$_7$, and E$_8$; we encountered the last of these in our discussion of heterotic string theory in Section \ref{ss.TOP} above.

\bibliographystyle{JHEP}
\bibliography{ref}
\end{document}